\documentclass[conference, a4paper]{IEEEtran}
\usepackage{braket}
\usepackage{amsmath}
\usepackage{dsfont}
\usepackage{graphicx}
\usepackage{cite}
\usepackage{mathtools}
\usepackage{amsfonts}
\usepackage{url}
\usepackage{bbm}
\usepackage{float}
\usepackage{stfloats}
\usepackage{tabularx}
\usepackage{xcolor}
\usepackage{cuted}
\usepackage{tikz}
\usepackage{balance}

\everypar{\looseness=-1}

\newcommand{\tr}{\text{Tr}}
\newcommand{\im}[1]{\text{Im}(#1)}
\newcommand{\re}[1]{\text{Re}(#1)}

\newcommand{\bigchi}{\makebox{\large\ensuremath{\chi}}}

\newcommand{\plus}{\mathord{\begin{tikzpicture}[baseline=0ex, line width=0.9, scale=0.06]
\draw (1,0) -- (1,2);
\draw (0,1) -- (2,1);
\end{tikzpicture}}}

\newcommand{\minus}{\mathord{\begin{tikzpicture}[baseline=0ex, line width=0.9, scale=0.06]
\draw (0,1) -- (2,1);
\end{tikzpicture}}}

\title{Protecting Classical-Quantum Signals in Free Space Optical Channels}

 \author{
 	 \IEEEauthorblockN{E. Villase\~nor,$^1$ M. S. Winnel,$^2$ T. C. Ralph,$^2$ R. Aguinaldo,$^3$ J. Green,$^3$ and R. Malaney.$^1$}\\
 	 \IEEEauthorblockA{
 	 ${}^1$ School of Electrical Engineering  \& Telecommunications,\\
 		 The University of New South Wales, Sydney, NSW 2052, Australia. \\
          ${}^2$ Centre for Quantum Computation and Communication Technology, \\
School of Mathematics and Physics, University of Queensland, St Lucia, Queensland 4072, Australia. \\
${}^3$ Northrop Grumman Mission Systems, San Diego,
California, USA.
      }
 }
\begin{document}

\maketitle

\begin{abstract}
Due to turbulence and tracking errors, free-space optical channels involving mobile transceivers are characterized by a signal's partial loss or complete erasure. 
This work presents an error correction protocol capable of protecting a signal passing through such channels by encoding it with an ancillary entangled bipartite state. 
Beyond its ability to offer protection under realistic channel conditions, novel to our protocol is its ability to encompass both classical and quantum information on the encoded signal. 
We show how, relative to non-encoded direct transmission,  the protocol can
improve the fidelity of transmitted coherent states over a wide range of losses and erasure probabilities. 
In addition, the use of ancillary non-Gaussian entangled bipartite states in the signal encoding 
is considered, showing how this can increase performance.
Finally, we briefly discuss the application of our protocol to the transmission of more complex input states, such as multi-mode entangled states.

\end{abstract}

%\vspace{-3cm}
\IEEEpeerreviewmaketitle

\section{Introduction}
\label{sec:introduction}
Building a global quantum network is a major technological challenge \cite{Pirandola2016}. Any quantum technology that relies on mature hardware may provide a practical path forward. 
Continuous Variable (CV) quantum communications, which can be implemented using off-the-shelf components, is one such technology \cite{Heim_2014,PhysRevLett.117.100503,PhysRevA.100.012325}. In this technology, quantum information is extracted using homodyne or heterodyne measurements - detection techniques previously developed to extract classical information encoded via coherent communications \cite{7174950, 9141375, 9484908}. Interestingly, CV quantum and coherent communications can coexist using the same signal \cite{Devetak2005, PhysRevLett.108.140501, Kumar_2015, PhysRevA.94.042340, Kumar2019}, and it is this interesting co-existence of classical and quantum information we study here. Our main contribution will be the introduction of a combined classical-quantum encoding protocol that accommodates a wide range of partial loss and erasure conditions in the channel. As we discuss later, such conditions are those anticipated for practical free-space optical channels, where at least one of the transceivers is mobile.

Optical free-space communications provide the basis for long-range communications, both classical \cite{4063386} and quantum \cite{PhysRevA.100.012325}, through horizontal channels on the ground \cite{Heim_2014, PhysRevA.100.012325} or more general channels to (or sometimes between) flying objects such as drones \cite{KUMAR2022100487}, aircraft \cite{Nauerth2013}, or satellites \cite{8439931, doi:10.1126/science.aan3211, Yin2020}.
However,  beam deformation and wandering induced by the turbulent atmosphere can have detrimental consequences \cite{PhysRevLett.117.090501, Wang_2018, Sebastian, sci3010004}. If at least one of the transceivers is mobile and untethered, then further misalignment of the optical beam introduced by transceiver motion (e.g., jitter) must also be considered \cite{20099, Song:17}. Any error correction protocol that can alleviate such combined effects would be useful for a pragmatic communication deployment.
 %In this work, we distinguish between loss, corresponding to photonic loss of the quantum signal, i.e., partial replacement of the signal by a vacuum state, and erasure, corresponding to complete replacement of the signal by a vacuum. In free-space communications, effects such as strong beam wandering coupled with other implementation deficiencies such as pointing and tracking errors give rise to erasures, uplink satellite communications being a prominent example \cite{andrews_book1, YI20132916}.
 
In the context of quantum information alone,
previous works have analyzed and experimentally demonstrated how error correction of the pure erasure channel (total loss in a channel) can be achieved via the introduction of ancillary modes transmitted through independent channels \cite{PhysRevLett.92.177903, PhysRevA.71.033814, doi:10.1080/09500340.2010.499047, Lassen2010, https://doi.org/10.48550/arxiv.2210.10230}.
Unlike these previous works, however, here we present a new protocol capable of error correction under more realistic conditions, as well as accommodating combined classical-quantum information encoding. The realistic channel conditions we consider encompass a  range of loss conditions across the independent channels  - not just simple erasures. As we shall see, careful monitoring of the channel losses via separate bright classical reference signals provides the feedback mechanism we need to protect a three-mode-encoded classical-quantum signal under our realistic channel conditions.

Our specific contributions in this work can be summarized as follows:  
(i)~We develop a three-channel error correction protocol capable of correcting any amount of loss on one mode of a three-mode quantum state while allowing some loss on the other  modes, demonstrating how to optimize signal recovery.
(ii)~We show that transmitted coherent states exhibit a  substantial improvement in fidelity over direct state transmission using one channel under our new protocol.  
(iii)~We demonstrate how transmission of classical information can be embedded into the protocol by adding a displacement operation, quantifying the displacement characteristics needed to achieve communications with a specific bit-error rate.
(iv)~We show how non-Gaussian states used as entangled ancillary states within the protocol can further increase the fidelity of transmitted states. 
(v)~We investigate whether the trends found for coherent states also apply to the distribution of entanglement by the protocol, discussing how care must be taken in interpreting the transmission of different input states.

This work is organized as follows.
In Section~\ref{sec:protocol}, the protocol is introduced. 
In  Section~\ref{sec:simulations}, simulation results of the state transmission via the protocol are presented.
The usefulness of non-Gaussian operations is presented in Section~\ref{sec:nonG}. 
Potential modifications to the protocol are discussed in Section~\ref{sec:discussion}, as are discussions on the use of alternate input states to the protocol. 
Our conclusions are drawn in Section~\ref{sec:conclusions}.

\section{The protocol}
\label{sec:protocol}
\begin{figure*}
   \centering
   \includegraphics[width=.99\textwidth ]{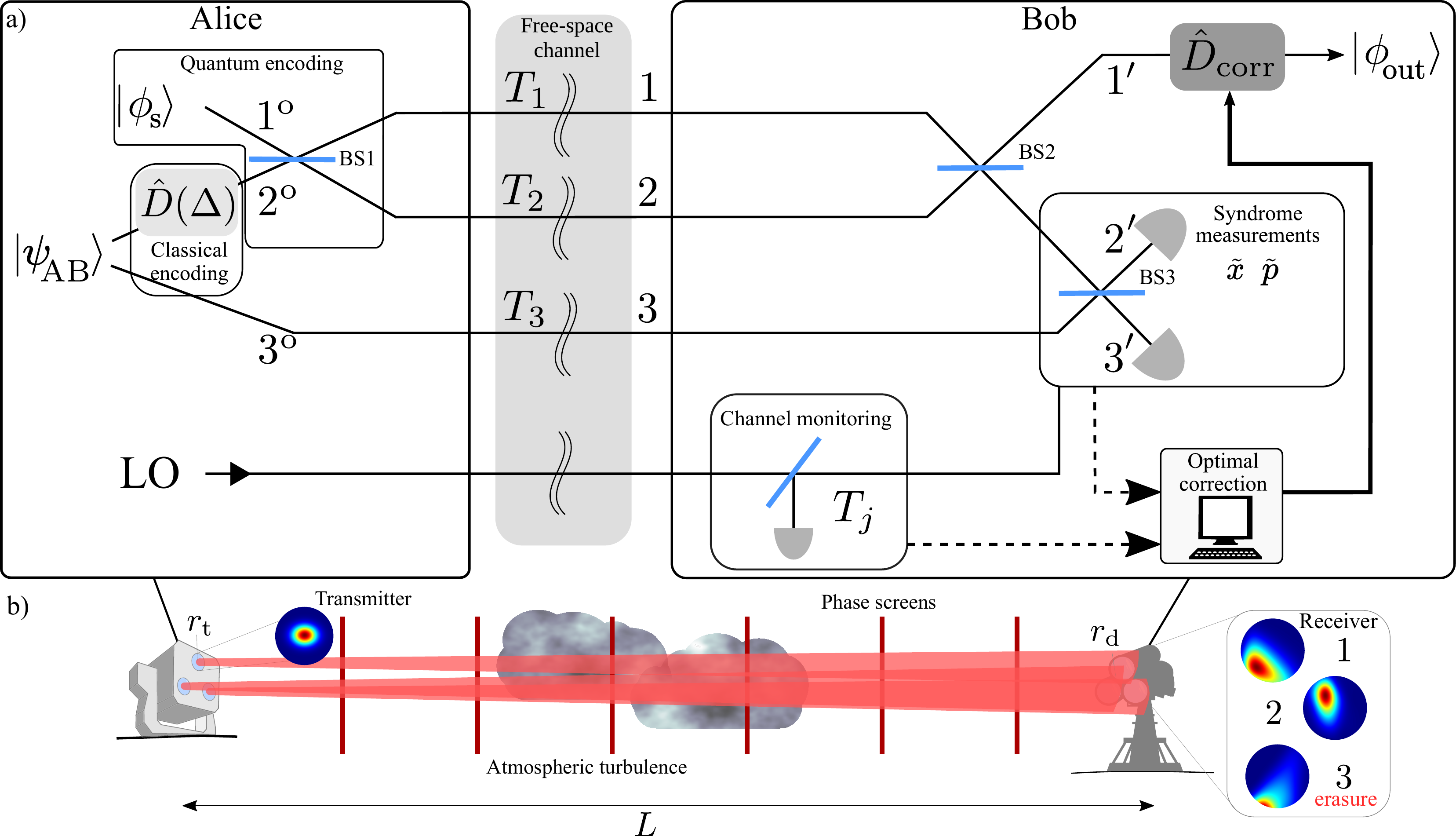}
   \caption{The protocol to transmit quantum and classical information (a). 
 The protocol protects a single-mode quantum state against erasures by combining it with an entangled bipartite state, $\ket{\psi_\mathrm{AB}}$. Classical information is encoded using the displacement operation applied in mode 2. Bob monitors the channels via the LOs. The syndrome measurement corresponds to a dual homodyne measurement. A correction is applied to mode $1'$ to recover the original quantum state using the syndrome result. 
 Dashed lines represent the transmission of classical information required to optimize the protocol. Turbulence simulations were obtained via phase screen simulations (b). Here, $r_t$ and $r_d$ represent aperture radii, and $L$ is the distance between transceivers.
 }
   \label{fig:qec}
 \end{figure*}
In Fig.~\ref{fig:qec}.a, 
our protocol is presented. The protocol consists of the concurrent transmission of classical and quantum information embedded into an error correction protocol. The error correction code is first described in the following, followed by the transmission of classical information via the protocol. 

The following steps are followed to perform error correction in the protocol. (i) Alice initially possesses the single-mode quantum state, $\ket{\phi_\mathrm{s}}$ the ``quantum signal" state to be error corrected in mode $1^\mathrm{o}$, alongside an entangled bipartite state, $\ket{\psi_\mathrm{AB}}$ in modes $2^\mathrm{o}$ and $3^\mathrm{o}$. (ii) The state $\ket{\phi_\mathrm{s}}$ is combined with mode $2^\mathrm{o}$ of the state $\ket{\psi_\mathrm{AB}}$ using a balanced Beam Splitter (BS), shown as BS1, to create a three-mode encoded state. (iii) The three modes of the encoded state are transmitted via the three fluctuating channels (acting independently on the three modes). The loss applied to each mode is characterized by the transmissivity coefficient $T_j$ with $j=1,2,3$. A Local Oscillator (LO) is transmitted from Alice to Bob through each channel (a bright classical beam of known amplitude multiplexed with each mode).  (iv) Bob receives the three modes, $1$, $2$, and $3$. (v) Bob obtains $T_j$  information on the channels by monitoring the three LOs. (vi) Bob uses a second BS (BS2), on modes $1$ and $2$ to decode the state. (vii) Applying a third BS (BS3), dual homodyne measurements are then performed   on modes $2'$ and $3'$ (referred to as the syndrome measurements $\tilde{x}$ and $\tilde{p}$). (viii) Bob uses his channel information, combined with the syndrome measurements, to determine the optimal corrective displacement, $\hat{D}_\mathrm{corr}$, that is applied to mode $1'$ to obtain the output state 
$\ket{\phi_\mathrm{out}}$ of the protocol.  If error correction was successful then, $\ket{\phi_\mathrm{out}}$ will deviate only marginally from $\ket{\phi_\mathrm{s}}$.

A critical component of the protocol is the classical computing required to apply the correction  $\hat{D}_\mathrm{corr}$. This classical computation can be thought of as an algorithm that takes as inputs the measured $T_j$ in each channel; the syndrome measurements, $\tilde{x}$ and $\tilde{p}$; and the (pre-measured) excess noise in each channel; and determines the optimal value of the gain, $g$, in $\hat{D}_\mathrm{corr}$  that will be applied to mode $1'$ so to optimize the fidelity between $\ket{\phi_\mathrm{out}}$ and $\ket{\phi_\mathrm{s}}$. The details of this optimization procedure are non-trivial and are explained in detail below. However, suffice it to say this procedure can be implemented in software \emph {a priori}. In the following, we will utilize one particular quantum information formalism to illustrate how the needed calculations within the software can be implemented.

The simulation results that appear later will be based on two steps: (i) the derivation of the analytical expressions for the transformations of quantum states in each step of  Fig~1.a; and  (ii) the numerical simulations to determine the probability distributions for the transmissivity of each channel (based on models of the phase screens indicated in Fig.~1.b) required as input to the analytical expressions.

\subsection{State transformation}
\label{sec:protocol:state}

A previous result \cite{MarianMarian} demonstrates that the outcome state of CV quantum teleportation can be elegantly computed using the  Characteristic Function (CF) formalism. In this work, we present a similar result.

For any $n$-mode quantum state, $\hat{\rho}$, its CF is defined as
\begin{align}
\bigchi(\lambda_1, \lambda_2, ..., \lambda_n) = \tr\left\{\hat{\rho} \hat{D}(\lambda_1)\hat{D}(\lambda_2)...\hat{D}(\lambda_n)\right\},
\end{align}
where $\lambda_i \in \mathbb{C}$, and $\hat{D}$ is the displacement operator,
\begin{align}
\hat{D}(\lambda_j) = e^{\lambda_j \hat{a}^{\dag}_j - \lambda_j^* \hat{a}_j},
\label{eq:displacement}
\end{align}
where $\hat{a}_j$ and $\hat{a}^{\dag}_j$ are respectively the annihilation and creation operators of mode $j$,  and ${}^*$ represents the complex conjugate.
Using the CF formalism, linear optics operations can be expressed by simply transforming the CF arguments while leaving the functions unchanged.
The effect of a loss-noise channel, with transmissivity $T_j$ and excess noise $\epsilon$ on a single-mode quantum state, corresponds to the transformation of its CF as,
\begin{align}
\bigchi\left(\lambda\right) \xrightarrow{\text{channel}}\bigchi\left(\sqrt{T_j}\lambda\right) \bigchi_{\ket{0}}\left(\sqrt{1 - T_j + \epsilon}\lambda\right),
\end{align}
where $\bigchi_{\ket{0}}$ is the CF of the vacuum state.

The full derivation of the CF of $\ket{\psi_mathrm{out}}$ is presented in Appendix~\ref{sec:ap:derivation}. Given the CF of the quantum signal, $\bigchi_\mathrm{s}(\lambda_1)$, and the CF of the entangled state, $\bigchi_\mathrm{AB}(\lambda_2,\lambda_3)$, the CF of the output state corresponds to,
\begin{align}
\bigchi_\mathrm{out}(\lambda)&= \bigchi_\mathrm{s} \left( (T_{\plus} + \tilde{g}T_{\minus}) \lambda \right) \times \\ \nonumber
&\bigchi_\mathrm{AB}\left( \left(T_{\minus} + \tilde{g} T_{\plus}\right)\lambda,  \sqrt{T_3}\tilde{g} \lambda^*\right) \nonumber \\ 
&\bigchi_{\ket{0}}\left(\sqrt{(1 + \epsilon) (2\tilde{g}^2 + 1) + 2g^2(1 - \eta^2) - T'} \lambda \right),
\label{eq:main}
\end{align}
with 
\begin{align}
T_{\plus} =  \frac{\sqrt{T_1} + \sqrt{T_2}}{2}; ~~~~ T_{\minus} =  \frac{\sqrt{T_1} - \sqrt{T_2}}{2}; \nonumber \\
T'= \frac{T_1}{2} (1+\tilde{g})^2 + \frac{T_2}{2} (1-\tilde{g})^2 + T_3 \tilde{g}^2.
\end{align}
 %The excess noise introduced by the channel is given by $\epsilon$, 
 Here, $\tilde{g}=g\eta$, with $\eta^2$ the efficiency of the homodyne measurements. The gain parameter when applying $\hat{D}_\mathrm{corr}$, $g$, corresponds to a free parameter Bob can select at will (see Appendix~\ref{sec:ap:derivation} for more detail). Error correction in the protocol works when Bob selects the appropriate value of $g$ based on the knowledge he obtains by monitoring the channels.
 By selecting the appropriate value of $g$, the output state can be made independent of the loss affecting one of the three modes in the encoded state.
The ideal scenario occurs when two of the three modes are unaffected by the loss. In this scenario, $g$ can always be selected to nullify the loss effects in the remaining mode. For more insight, some idealized scenarios with specific $g$ values are discussed in Appendix~\ref{sec:ap:examples}.

\emph{Transmission of classical information}:
The only additional requirements for transmitting classical information are an agreed digital modulation scheme between Alice and Bob and an extra displacement operation $\hat{D}(\Delta)$. During the first step in the error correction, the operation $\hat{D}(\Delta)$ is used by Alice to encode a symbol on mode $2^\mathrm{o}$ of $\ket{\psi}_\mathrm{AB}$ immediately after it has been prepared. The symbol corresponds to one or multiple classical bits following a predetermined digital modulation scheme. 
The symbol is then recovered by Bob automatically from the values of the syndrome measurement results, $\tilde{x}$ and $\tilde{p}$.
Note, we adopt $\hbar=2$; and $\Delta$ , $\tilde{x}$ and $\tilde{p}$ are dimensionless (the variance of the vacuum noise is one).

\subsection{Transmissivity calculations}
\label{sec:protocol:trans}

The technique used here to model turbulence via phase screen simulations has been detailed extensively in our previous works, with experimental validation over a horizontal free-space channel of 1.5~km \cite{9348086}. The simulation methodology in the present work follows our previous works; details can be seen in \cite{9348086, 9463774}.

\begin{figure}
   \centering
   \includegraphics[width=.5\textwidth]{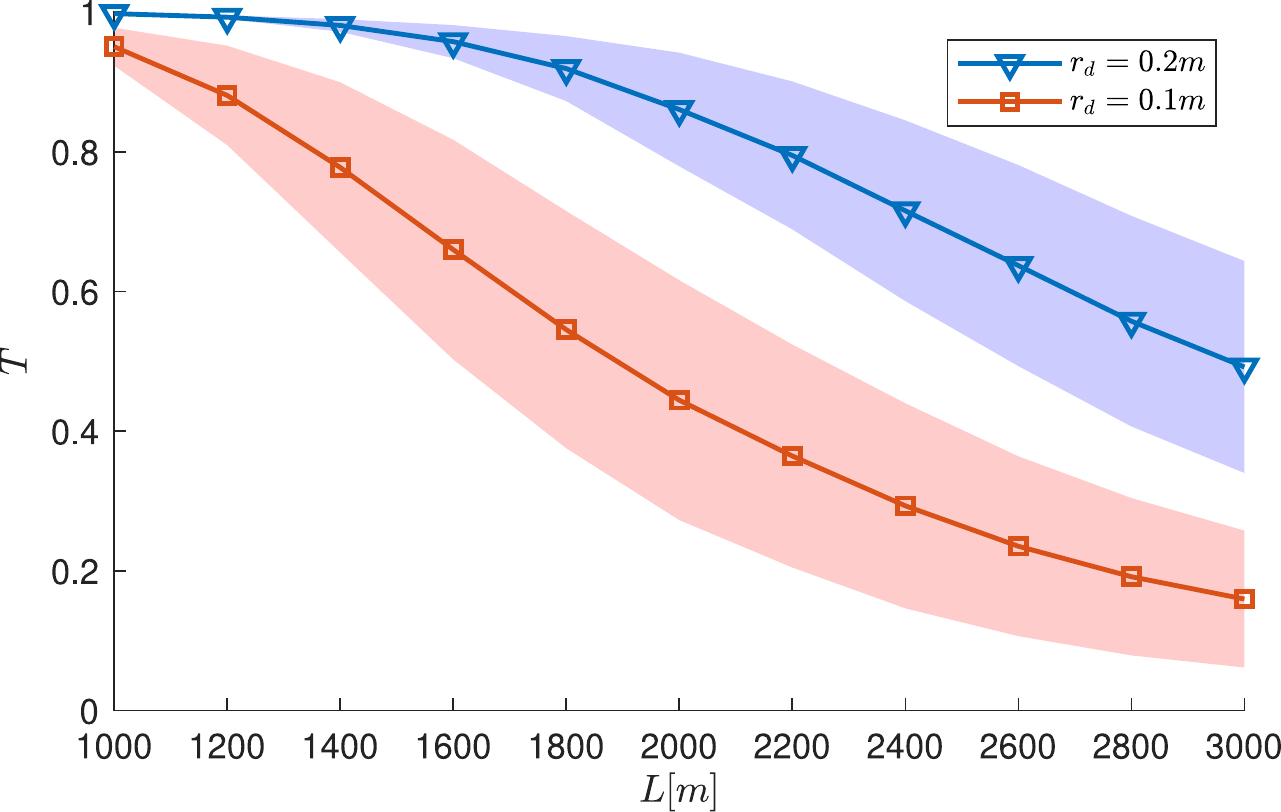}
   \caption{Transmissivities obtained from the model of the free-space channel following the parameters in Table I. The shaded area represents the standard deviation of the curves.}
	
	\label{fig:pdf}
 \end{figure}

Our phase screen simulations allow us to find the Probability Density Function (PDF) of the transmissivity in the channel for any given communications setup.
Henceforth, we refer to this PDF as  $\mathcal{P}$. The dominant parameters of a communication setup are the propagation distance, $L$, the wavelength of the light used, $\lambda$, the beam waist at the transmitter, $r_\mathrm{t}$, and the size of the receiving aperture, $r_\mathrm{d}$.
These parameters will be fixed to be values shown in Table~I (with the exception of $L$ and $r_\mathrm{d}$).  For focus, we assume a horizontal channel in the results we show here; however, other configurations can be easily accommodated via small changes in the simulations \cite{9348086}.
A total of 10 uniformly distributed phase screens are used, and 
the grid size of complex numbers is $1500 {\times} 1500$. A typical PDF from our calculations can be seen in Fig.~\ref{fig:pdf}.

Beyond turbulence, other real-world system issues can affect the channel transmissivity. Most notable are the pointing and tracking errors between the transmitter and receiver that create a jitter and/or deterministic offsets of the beam direction \cite{20099, Song:17}. In our channel model, we will account for these effects by the addition of erasures into the PDF of the transmissivity. That is, for each optical beam sent from the transmitter to the receiver, we consider that there is a probability $p_\mathrm{e}$ of it entirely missing the receiver (corresponding to $T_j=0$). In other words, for a given communications setup (e.g., the transceiver apertures, propagation distance, and wavelength) with corresponding PDF  $\mathcal{P}$, and an erasure probability, $p_\mathrm{e}$, the values of $T_j$ are now given by the following modified PDF,
\begin{align} \label{eq:Tpdf}
    T_j = \begin{cases} 
      T_j \sim \mathcal{P} & \mathrm{with \ probability \ }(1-p_e) \\
      0 & \mathrm{with \ probability \ }p_e,
   \end{cases}
\end{align}
where $X\sim Y$ indicates that a random variable $X$ is drawn from distribution $Y$. 
This channel model represents a good approximation to the 
real-world channel of interest to us - a channel where erasure probability is high\footnote{In channels where  $\mathcal{P}$ indicates a small probability for $T\sim 0$, and where $p_e\sim0$, we find that no improvement from the protocol is likely - the transmission may as well be just direct. If \emph {a priori} known that $p_e=0$,  we have a Gaussian channel, and the introduction of non-Gaussian states or measurements into the protocol is needed if any advantage is to be forthcoming.}. Also, the extension of the channel model via a new probability, $p_e$, 
to parameterize the additional loss caused by transceiver movement/offsets  allows us to compare 
better with previous related work.

\begin{table}
\centering
\caption{Free-space communications parameters}
\begin{tabular}[t]{ccccc}
$\lambda$ [nm] &$r_t$ [cm] & $C_n^2$($\text{[m]}^{-2/3}$) & $l_0$ [mm] & $L_0$ [m]\\ 
\hline
1550& 2.5 & $2.47\times 10^{-13}$ & 7.5 & 1.57 \\
\hline
\label{tab:values}
\end{tabular}
\end{table}

\section{Simulations}
\label{sec:simulations} 
Here, we first outline the CFs of the main quantum resources we use and make clear how we determine the fidelity - the key metric we use. We then discuss the performance of the quantum and classical communication under the protocol, detailing how the optimization of $g$ is determined for each simulation run.
\subsection{Quantum resources}
The quantum signal is a coherent state $\ket{\alpha}$, whose CF is,
\begin{align}
  \bigchi_{\ket{\alpha}}(\lambda) = \exp \left[-\frac{|\lambda|^2}{2} + (\lambda \alpha^* - \lambda^* \alpha) \right].
\end{align}
The two-mode squeezed vacuum (TMSV) state $\ket{\psi_\mathrm{AB}}$ used in the encoding has a CF,
\begin{align}\label{eq:cftmsv}
\bigchi_\mathrm{TMSV}(\lambda_{\mathrm{A}}, \lambda_{\mathrm{B}}) = \exp \left[ -\frac{1}{2}\left(|\lambda_\mathrm{A}'|^2 + |\lambda_\mathrm{B}'|^2 \right) \right].
\end{align}
In deriving this, the following
 transformation has been used\cite{DellAnno1}, 
\begin{align}
\label{eq:Bogo}
&\hat{S}_\mathrm{AB}(\varrho) \hat{D}(\lambda_\mathrm{A}) \hat{D}(\lambda_\mathrm{B}) \hat{S}^\dag_\mathrm{AB}(\varrho) = \hat{D}(\lambda_\mathrm{A}') \hat{D}(\lambda_\mathrm{B}'),  \\
&\lambda_j' = \cosh(r)\lambda_j + e^{i\phi} \sinh(r) \lambda_k^* ~~~~ j,k = \{\mathrm{A,B}\}; ~~ j\neq k, \nonumber
\end{align}
with $\hat{S}_{\mathrm{AB}}$ the two-mode squeezing operator, and with $\varrho=re^{i \phi}$. Without loss of generality, the value $\phi=\pi$ is set.
% Here (and throughout this work), we assume $\hbar=2$.
 %The squeezing magnitude $r$ is directly proportional to the entanglement of the TMSV state.
Finally, the CF of the vacuum state $\ket{0}$, that appears in Eq.~\ref{eq:main} is,
\begin{align}
\bigchi_{\ket{0}}(\lambda) = \exp \left[ -\frac{|\lambda|^2}{2} \right].
\end{align}

Fidelity is used as the metric of the effectiveness of the protocol.
The fidelity represents the closeness between the quantum signal $\ket{\psi_s}$ and the output state $\ket{\psi_\mathrm{out}}$. 
In the CF formalism, the fidelity is computed as,
\begin{align}
\mathcal{F}_\alpha = \frac{1}{\pi} \int d^2 \lambda \bigchi_\mathrm{s}(\lambda) \bigchi_\mathrm{out}(-\lambda).
\label{eq:fidelity}
\end{align}

The fidelity as defined in Eq.~\ref{eq:fidelity} will depend on each input state's value $\alpha$. Therefore, the mean fidelity over an ensemble of coherent states must be considered. The following Gaussian distribution specifies the ensemble,
\begin{align}\label{eq:probcoherent}
P_\alpha(\alpha) = \frac{1}{\sigma_\alpha \pi} \exp\left[-\frac{|\alpha|^2}{\sigma_\alpha}\right],
\end{align}
with $\sigma_\alpha$ the variance.
Therefore, the fidelity over the ensemble of coherent states will be used, defined as
\begin{align}\label{eq:avefidelity}
\mathcal{F}  = \int d{\alpha}^2 P_\alpha(\alpha) \mathcal{F}_\alpha.
\end{align}

Ideally, we would like to compute the fidelity considering a uniform distribution of coherent states (corresponding to $\sigma_\alpha \to \infty$); however, this would be analytically intractable. An approximation to a uninform distribution can be obtained by setting a large enough variance, $\sigma_\alpha = 10$ (a value we adopt through this work).
%\footnote{For example, between $\sigma_\alpha = 10$  and $\sigma_\alpha = 20$ the differences in fidelity are of the order of $10^{-2}.} 
For reference, the so-called {\it classical limit}, $\mathcal{F}_\mathrm{class}=0.52$ sets the baseline where fidelities of transmission below this value could be obtained using purely classical communications\cite{PhysRevLett.94.150503}.

Using Eq.~\ref{eq:main} with Eq.~\ref{eq:avefidelity} it becomes possible to find an analytical expression of the fidelity for the ensemble of coherent states. This expression is provided in Appendix~\ref{sec:ap:expressions}, Eq.~\ref{eq:Fprotocol}. Additionally, the fidelity for direct transmission for the ensemble of coherent states, $\mathcal{F}^{\mathrm{dir}}$, is also presented in Appendix~\ref{sec:ap:expressions}, Eq.~\ref{eq:Fdirect}.

\subsection{Transmission of quantum information}
\label{sec:simulations:quantum}
\begin{figure}
  \centering
  \includegraphics[width=.5\textwidth]{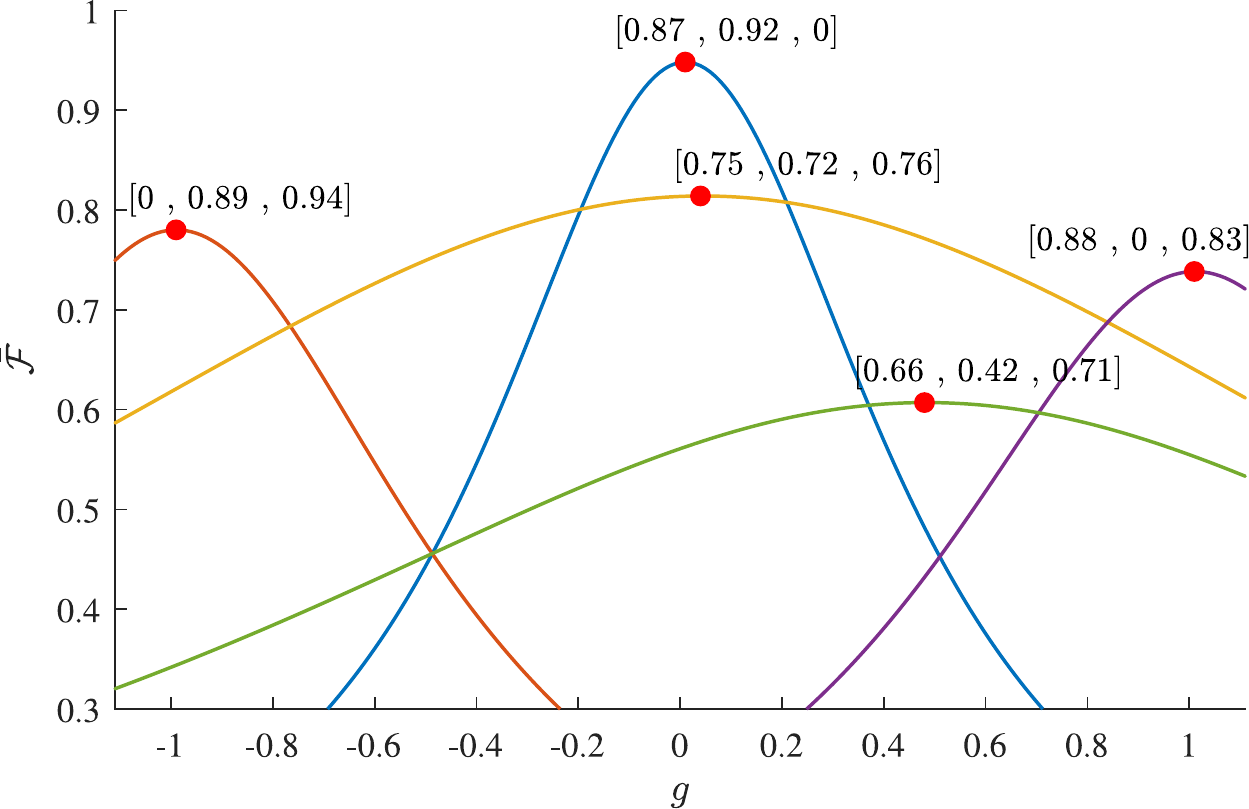}
  \caption{Fidelity as a function of $g$, each curve corresponds to a specific set of values of $[T_1, T_2, T_3]$. The red dots mark the value of $g=g_\mathrm{opt}$ corresponding to the maximum fidelity, and each label denotes the values of the set $[T_1, T_2, T_3]$ corresponding to the adjacent curve. }
  \label{fig:gs}
 \end{figure}

In the following calculations, a homodyne measurement efficiency of $\eta^2=0.9$ is set unless specified otherwise. Fluctuations of the free-space channel produce several effects contributing to the excess noise \cite{Wang_2018, Sebastian}. We have modeled the excess noise in our simulation as  $\epsilon= \epsilon_\mathrm{ph} + \epsilon_\mathrm{det}$. Here,  $\epsilon_\mathrm{ph}$ is related to noise introduced by the displacement used in the encoding  (see  \cite{PhysRevA.94.042340, Kumar2019}) and will be modified by the transmissivity of the channel. The other term,  $\epsilon_\mathrm{det}$  is detector excess noise, which we set to $\epsilon_\mathrm{det}=0.013$  \cite{Sebastian}. As an example, for $T=0.5$,  we have $\epsilon=0.018$. 
When computing the fidelity, every protocol realization involves a different sample of the values $T_j$. Thus, the parameter $g$ must be optimized for each protocol realization, as shown in Fig.~\ref{fig:gs}.
The following procedure was followed to compute the mean fidelities. First, using the numerical simulations described in Section~\ref{sec:protocol:trans}, 30000 samples of $T_j$ were obtained for a given 
propagation distance. Next, fidelity was obtained for a specific realization of the channels using three sampled values of $T_j$ in Eq.~\ref{eq:Fprotocol}.

All possible transmissivity occurrences must be accounted for to compare the protocol's effectiveness to direct transmission. 
For direct transmission, the total fidelity is calculated as 
\begin{align}
\mathcal{F}^\mathrm{dir}_\mathrm{total} = (1- p_\mathrm{e}) \mathcal{F}^\mathrm{dir} + 
p_\mathrm{e}\mathcal{F}_{\ket{0}}.
\end{align}

 \begin{figure}
   \centering
   \includegraphics[width=.5\textwidth]{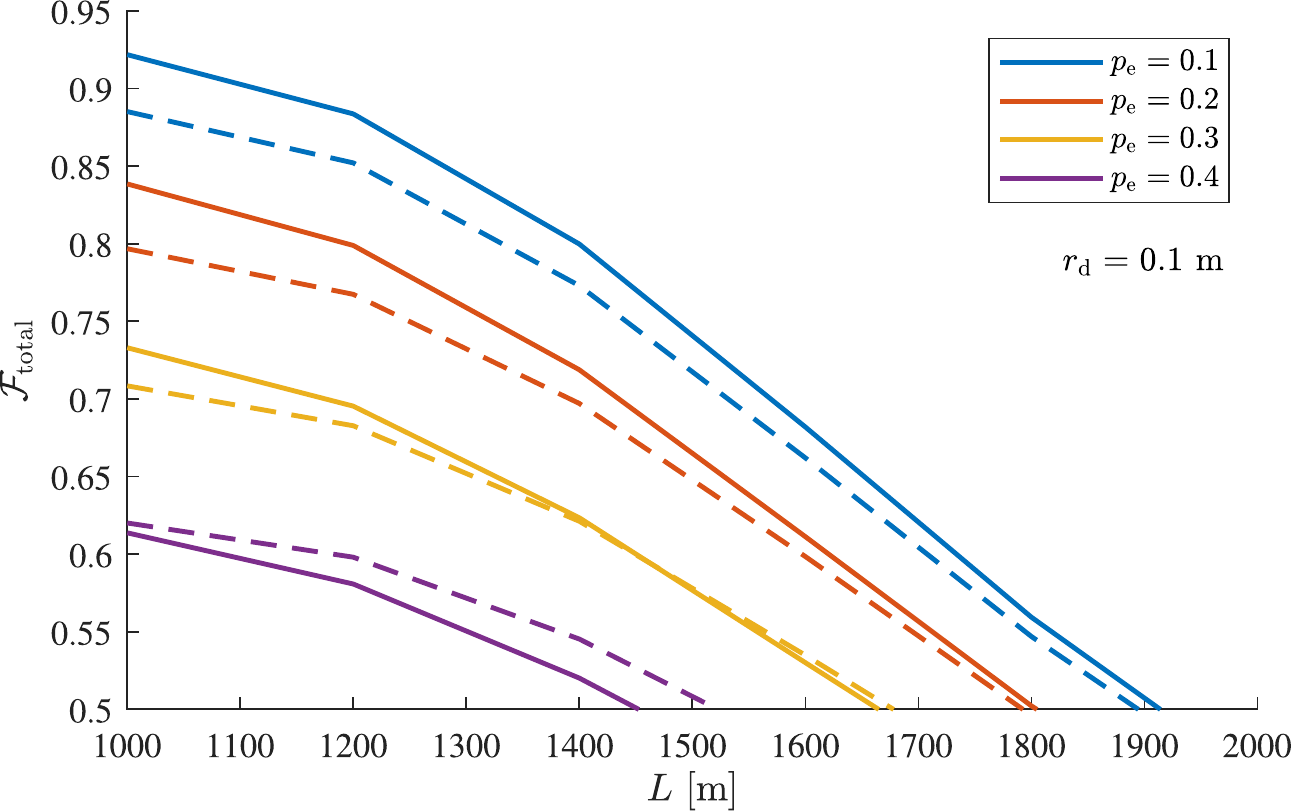}
   \includegraphics[width=.5\textwidth]{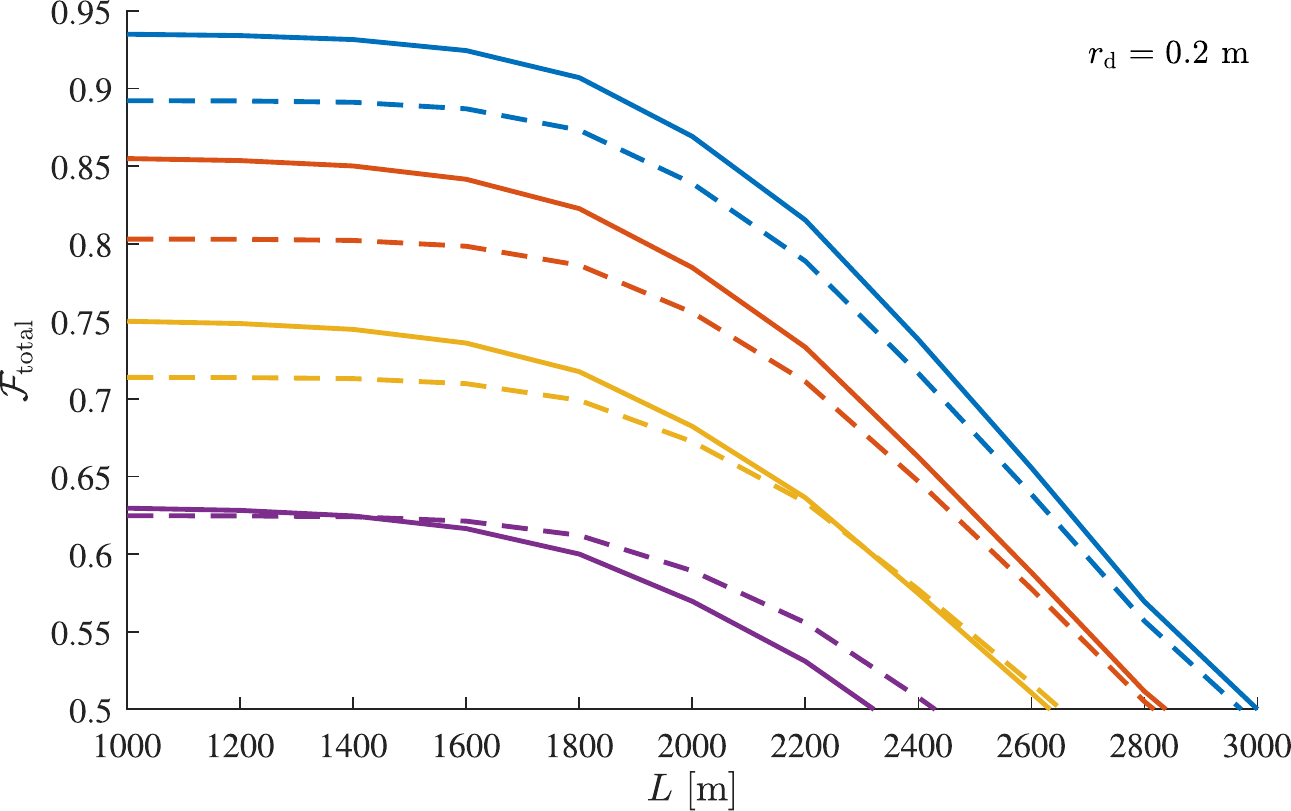}
   \caption{Total fidelity for different values of erasure rate. Solid lines represent transmission via the protocol, and dashed lines correspond to direct transmission. The ancilla entangled state uses a squeezing value of $r=10$ dB.}
   \label{fig:fid_total}
 \end{figure}

The results comparing the mean total fidelity between the protocol and direct transmission are shown in Fig.~\ref{fig:fid_total}. The results indicate that the protocol improves the transmission fidelity for values of $p_\mathrm{e}$ up to 0.35. The advantage provided by the protocol decreases as the distance increases. For values, $p_\mathrm{e}{>}0.35$, it is observed that the protocol does not provide any advantage for the $r_d$ values shown.

\subsection{Transfer of classical information}
\label{sec:simulations:classical}
Now the error rates in transmitting classical information via the protocol are analyzed.
An error during the transmission of classical information corresponds to Bob misidentifying the symbol sent initially by Alice. The excess noise added to the system increases the variance of the syndrome measurements. Thus, if the noise is too high, there is a probability that the syndrome measurements appear on a different partition from the one Alice meant to encode. The error probability can be directly quantified by the ratio between the variance of the syndrome measurements, $\sigma_s$, and the magnitude of the displacement used during the classical encoding, $|\Delta|$.
For a given value of $|\Delta|$, a larger $\sigma_s$ means a higher Bit Error Rate (BER). Alternatively, a more significant displacement can also be used to increase the distance between partitions and offset the effects of $\sigma_s$, increasing the Signal-to-Noise Ratio (SNR).

If all the states used during the protocol are Gaussian, then the measurement results will follow a Gaussian distribution \cite{RevModPhys.84.621}.
Solving the integration in Eq.~\ref{eq:prob_dist} in Appendix~\ref{sec:ap:derivation}, we see that the syndrome measurements have the mean values given by  
 \begin{align} \label{eq:mu_class}
\mu_{\tilde{x}} =  \frac{1}{\sqrt{2}}\eta T_{\plus} \re{\Delta} \nonumber \\ 
 \mu_{\tilde{p}} = \frac{1}{\sqrt{2}}\eta T_{\plus} \im{\Delta}.
\end{align}
The variance $\sigma_s$ depends on the specific values $T_j$, the amount of squeezing $r$, and the size of the ensemble of states being transmitted using the protocol $\sigma_\alpha$. The expression of $\sigma_s$ is presented in Appendix~\ref{sec:ap:snr}.

Once we have determined the moments of the syndrome measurements, we can compute the BER. For simplicity, we assume the classical information consists of single bits encoded in Binary Phase-Shift Keying  \cite{7174950}. In this case, the BER of the protocol will be,
\begin{align}
\mathrm{BER}=\frac{1}{2} \mathrm{erfc}\left(\frac{\eta T_{\plus} |\Delta|}{\sqrt{2\sigma_s^2}}\right).
\end{align}

\begin{figure}
   \centering
   \includegraphics[width=.5\textwidth]{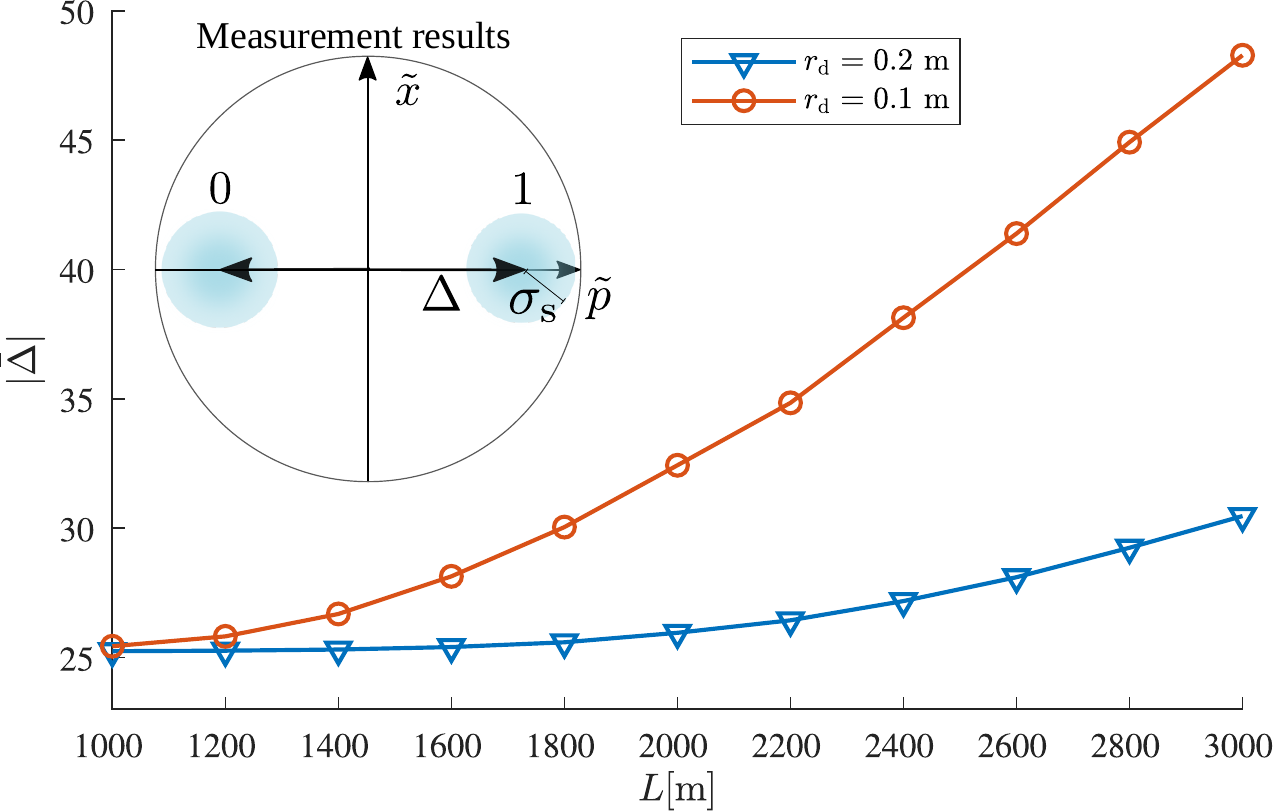}
   \caption{Mean magnitude of the displacement, in shot-noise units, required to keep the error rate below $10^{-9}$ when the transmissivity is zero on mode $1$. A squeezing value $r=10$ dB is considered in the TMSV state. 
     }
   \label{fig:disp}
 \end{figure}

A value of interest is the size of the displacement required to maintain the BER to a tolerable level. For a BER of $10^{-9}$, we see in Fig.~\ref{fig:disp} the mean values of $|\Delta|$ required for the special case when $T_1=0$ in mode $1$ (or similarly mode $2$). 
If instead mode $3$ possesses $T_3=0$, the displacement requirements are approximately half of the values shown.

\section{Enhancement via non-Gaussian operations}
\label{sec:nonG}
Non-Gaussian states represent a key element in CV quantum information, and there has been great interest in their use to enhance quantum communications protocols \cite{MingjianPAPS, 9024548, PhysRevA.91.063832, Lim2019, ScissorsQKD, PhysRevA.102.052425}. Particularly, in CV quantum teleportation, certain non-Gaussian operations have shown to enhance the fidelities of transmitted states \cite{nonGaussianTeleportation, DellAnno1, 9685767, DAT2022168744}. Thus, it is natural to ask if the same holds for this protocol.

\begin{figure}
\centering
\includegraphics[width=.3\textwidth]{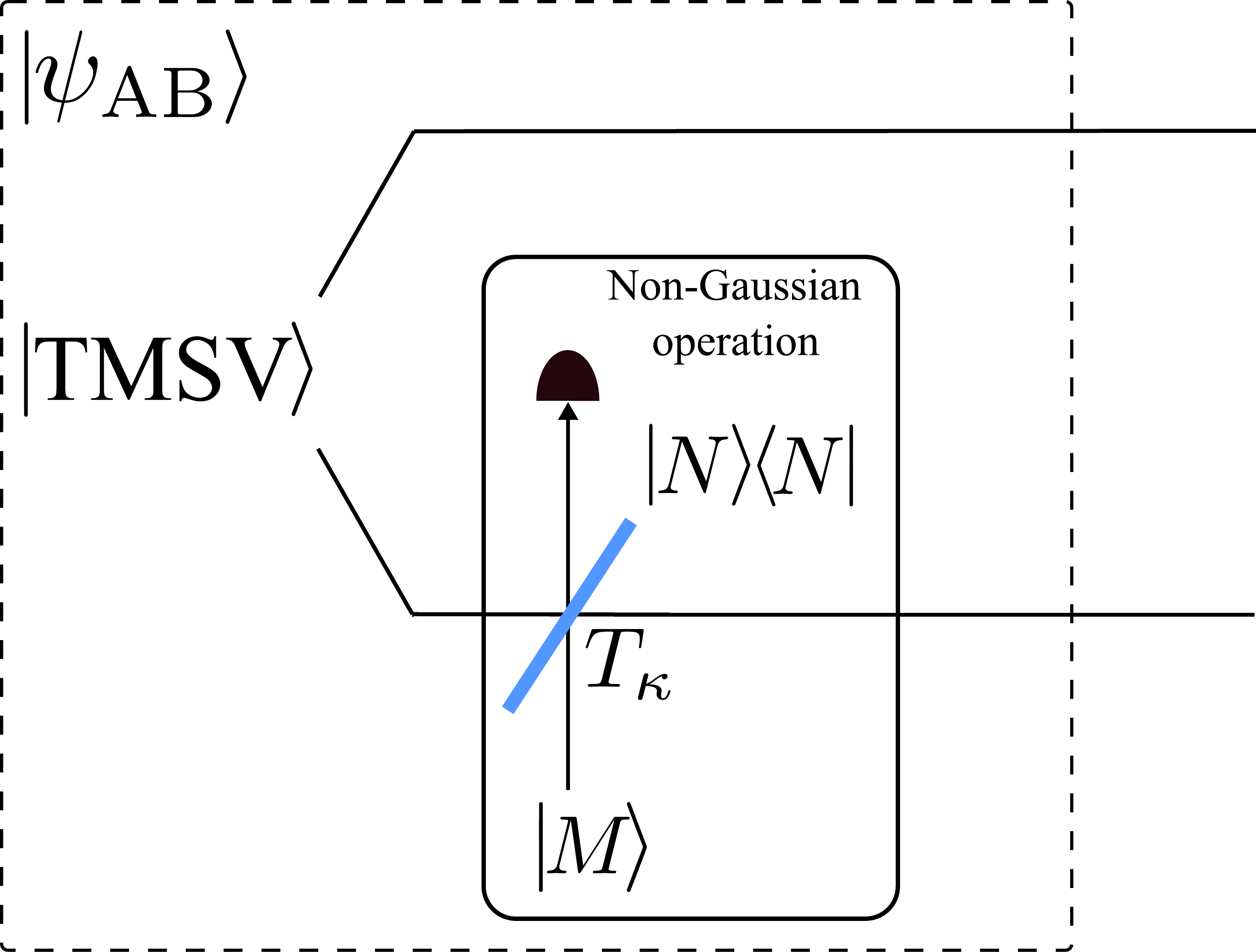}
\caption{Non-Gaussian operations used to generate a non-Gaussian resource state to be used in the teleportation channel. The operations are applied on a single mode of a TMSV state.}
\label{fig:operations}
\end{figure}

One viable experimental way to construct non-Gaussian states 
shown in Fig.~\ref{fig:operations}, where a mode of a TMSV is combined with a BS with transmissivity $T_\kappa$ with a $M$ photon Fock state, and a photon discriminating detector is used to detect $N$ photons. In the following, we consider a Photon Subtraction (PS), setting $M=0, N=1$; Photon Addition (PA) with $M=1, N=0$; and Photon Catalysis (PC) setting $M=1, N=1$.
We also consider the successive application of the PS and PA operations,  PS-PA (order reversed in PA-PS). 
All the operations are applied in mode $3^\mathrm{o}$ of the initial TMSV state as the first step in our protocol. 

Besides applying non-Gaussian operations on a TMSV, an additional non-Gaussian entangled state is considered. This state is prepared at the start of the protocol by the application of the two-mode squeezing operation to a Bell state, the state known as the {\it Squeezed Bell} (SB) state, given by
\begin{align}
\ket{\psi_\mathrm{SB}} &=(\cos^2(\vartheta) + \sin^2(\vartheta))^{-1/2} \nonumber \\ & \times  \hat{S}_\mathrm{AB}(\varrho) \Big[\cos(\vartheta) \ket{00} + \sin(\vartheta)\ket{11} \Big].
\end{align}

The implementation of the non-Gaussian states is achieved by replacing $\ket{\psi_\mathrm{AB}}$ in Fig.~\ref{fig:qec}  with the corresponding non-Gaussian state. The CF of each non-Gaussian state must be computed first and used in Eq.~\ref{eq:main}, where the non-Gaussian CF replaces $\bigchi_\mathrm{AB}$. 
The CF of the states used in this work have been calculated in previous works \cite{9685767}, and are listed in Appendix~\ref{sec:ap:nonG} for completeness.

In the CF of each non-Gaussian state, a free parameter exists corresponding to the transmissivity of the beam-splitter involved in each operation. This free parameter is optimized to maximize the fidelity of transmission. In the case of the SB state, the $\vartheta$ free parameter is optimized. For the PS-PA and PA-PS, the exact value of $T_\kappa$ is used in both successive non-Gaussian operations. Finally, it is assumed that a quantum memory is available at the transmitter, such that the non-Gaussian states can be prepared and stored to be used on demand during the protocol\footnote{Non-Gaussian operations have a non-unity success probability. Without a quantum memory, this success probability would need to be accounted for in the resulting fidelities.}.

\begin{figure}
   \centering
   \includegraphics[width=.45\textwidth]{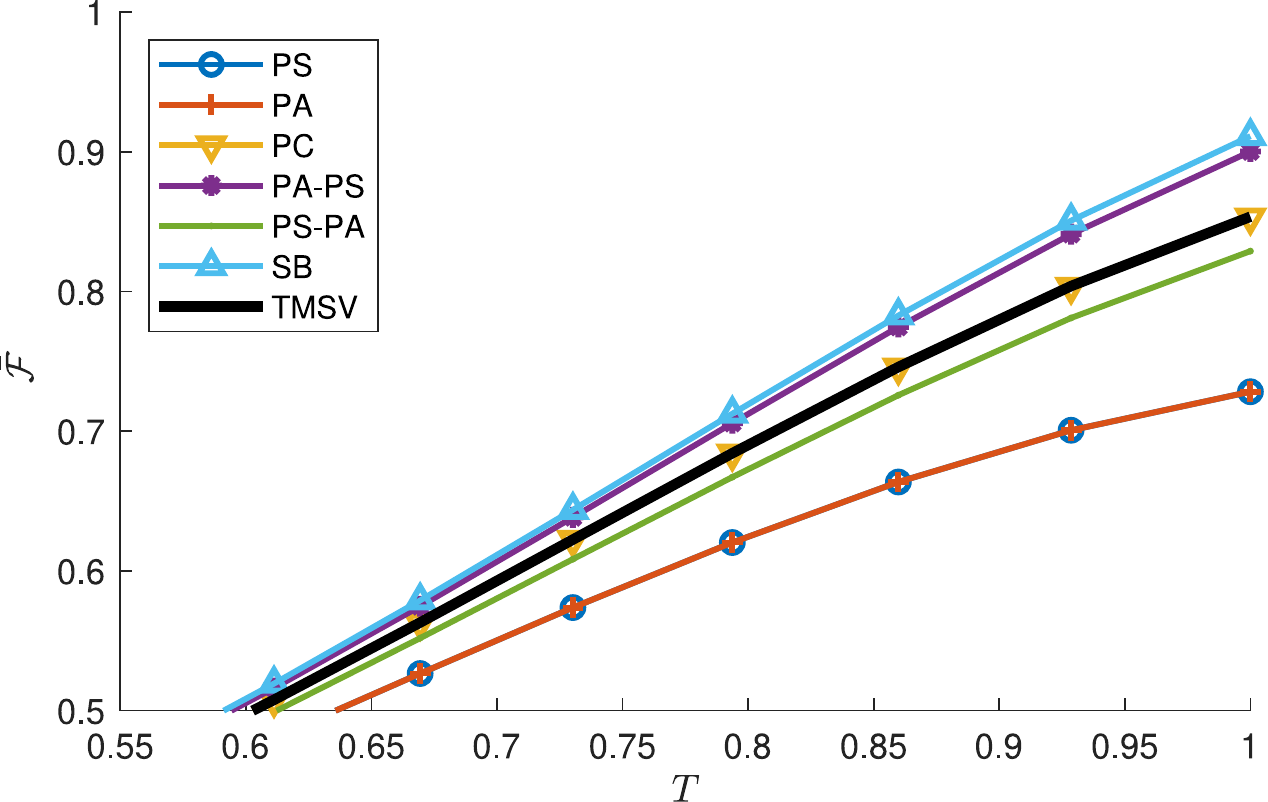}

   \caption{Fidelities obtained for a loss channel for the different non-Gaussian states. In each case, the squeezing of the TMSV state before the non-Gaussian operation is fixed at a value $r=4.7$ dB. }
   \label{fig:nonG}
 \end{figure}

The results are presented in Fig.~\ref{fig:nonG}. To simplify the analysis, we evaluate the fidelities for a simplified channel, where  $\mathcal{P}= \delta(T_j-T')$ in Eq.~\ref{eq:Tpdf}, the excess noise is set as $\epsilon = 0$, and $\eta=1$.
Of all the non-Gaussian resources used, only two states represent an enhancement over the TMSV squeezed state: the PA-PS and the SB states. Moreover, the gain in fidelity is only observed when the squeezing in the entangled states is low, around $r=4.7$ dB. As the squeezing increases, the
fidelities obtained by the non-Gaussian states get closer to the ones obtained by the TMSV state, for a value of $r=10$ dB, the difference in fidelity is of the order of $10^{-2}$.
We point out that there is an agreement between the two non-Gaussian resources that represent an enhancement between the protocol presented here and CV quantum teleportation \cite{9685767}. Moreover, when mode $3$ is erased, the fidelities obtained using the non-Gaussian states are equal to the ones obtained when the TMSV state is used.

Now, we address how the use of non-Gaussian states affects classical encoding. When non-Gaussian states are used, the syndrome measurement results follow a non-Gaussian PDF. Numerical integration over this non-Gaussian PDF indicates that the values of $|\Delta|$ required to keep low BER are considerably higher when the non-Gaussian states are used. Approximately $80\%$ larger $|\Delta|$ is required in the SB and PA-PS states when mode $1$ is erased\footnote{A case where the application of the PA-PS operation is applied to mode $2^\mathrm{o}$ instead of mode $3^\mathrm{o}$ during the preparation of the non-Gaussian state was also tested. The results showed no difference in the displacement requirements between the two cases.}.

\section{Discussion}
\label{sec:discussion}
Here, we discuss our new results in relation to previous similar work, some alternative encoding schemes, and some additional input states.

\subsection{Relation to other work}
Other works 
\cite{PhysRevLett.92.177903, PhysRevA.71.033814, doi:10.1080/09500340.2010.499047, Lassen2010, https://doi.org/10.48550/arxiv.2210.10230} have previously looked at similar encoding schemes. In \cite{PhysRevLett.92.177903, PhysRevA.71.033814}, a three-mode encoding scheme was studied in the context of a quantum state sharing scheme. Although not aimed at the communication scenario, the nature of this scheme was similar to the encoding required for an erasure channel in the communication context. In \cite{Lassen2010}, a  four-mode encoding scheme used to transfer two coherent input states over an erasure channel was investigated. In \cite{https://doi.org/10.48550/arxiv.2210.10230},  a three-mode encoding scheme similar to that used in the context of state sharing was analyzed for the communication scenario, with extensions to include non-Gaussian operations. This latter work was again only in the scenario of an erasure channel. Channel errors beyond a simple erasure were considered in \cite{doi:10.1080/09500340.2010.499047} - namely, a displacement error.

The results presented here represent outcomes different from all these previous works in several aspects. First, we consider a  channel model more appropriate to that anticipated for free-space atmospheric communications in which one transceiver is untethered. Simulations of this channel are then used to analyze a modified protocol deployed over a three-mode channel. The key modification in the protocol is an optimization phase in which the transmissivity measured for each channel is used as an input. This allows for encoding and decoding in the more general case where transmissivities in all channels are non-zero. Importantly, our protocol leads to improved results relative to the situation where any loss is simply considered a complete erasure.  Second, we have included additional displacements within the encoding phase to include classical signaling. Third, we have investigated non-Gaussian operations within the modified protocol.

\subsection{Alternative  encodings and post selection}
Thus far, we have considered combined classical-quantum communication embedded in the same signal. However, other possibilities exist. For example, the reference pulses could embed classical information – simple on-off keying being one scenario. In this scenario, the lack of any LO sent in a channel could indicate a zero and its presence, a one. The quantum signal, in this case, would not need an additional displacement, but the trade-off would be that the quantum information rate would be reduced. Exactly how much reduction would occur would depend on the coherence time of the channel and the pulse rate of the source, but for anticipated timescales of order 1ms and available pulse rates of 100MHz, this reduction would be minimal (since the channel is stable for thousands of sent LOs - these can be re-used). This trade-off of a reduced quantum information rate would have the benefit of a reduction in the excess noise of ${\approx} 0.02$ and would yield an increase of fidelity (${\approx}0.05$).  
An added benefit would be a reduction of the complexity of deployment (one less displacement operation) and potential power savings (fewer LOs sent). On-off keying could also be applied to the quantum signals (with a vacuum detection being mapped to a zero). However, such schemes would need a modified digital modulation scheme  (to avoid `on' signals which have small displacements).

Our results presented here have utilized an averaging over all channel conditions. In practice, post-selection could be utilized to select the channels such that, at most only one presents an erasure. Again, a trade-off in performance vs. throughput would be in play here. Post-selection would be most beneficial under high erasure error rates. For example, post-selecting to allow at most a single erasure at an error rate $p_\mathrm{e}=0.3$ would have the benefit of the fidelity being increased by ${\approx}0.15$, at a reduction in throughput of 0.8.

\subsection{Entanglement distribution}
We are also interested in evaluating the protocol's effectiveness in distributing entangled states.
Considering an input TMSV state, $\ket{\psi^\mathrm{in}_\mathrm{MN}}$ with squeezing $r_\mathrm{MN}$, we can evaluate the fidelity with the state obtained after mode $\mathrm{M}$ has been transmitted via the protocol. Although not shown, our results show that there are scenarios where an advantage over direct transmission can be found similar to before.
%This calculation is straightforward and returns an exact value \cite{PhysRevLett.115.260501}. 
For equal squeezing values in the states $\ket{\psi_\mathrm{AB}}$ and $\ket{\psi^\mathrm{in}_\mathrm{MN}}$, we observe the use of the protocol does present an advantage over direct transmission.

Ideally, for entangled input states, we would like to test the protocol's effectiveness using an application of quantum communications, such as entanglement distribution.  However, using only fidelity can be an incomplete metric of the transmission of quantum information, especially for entangled states \cite{PhysRevResearch.4.023066}. 
Entanglement distribution is perhaps better measured by computing the \emph{Reverse Coherent Information} (RCI) of the channel, as it
represents a lower bound on the distillable entanglement \cite{PhysRevLett.102.210501}.
The RCI is defined as,
\begin{align}
    \mathcal{R}= S(\rho_\mathrm{N}) - S(\rho_\mathrm{MN}),
\end{align}
with $S(\cdot)$ the Von Neumann entropy, and $\rho_\mathrm{MN}$ a maximally entangled TMSV state ($r_\mathrm{MN} \to \infty$), where mode $\mathrm{M}$ has been transmitted via the channel using the protocol.

For our protocol, an upper bound on the RCI can be found using the fact that the entropy is  concave,
\begin{align}
    S(\rho_\mathrm{MN}) \geq \sum_k p_k S(\rho_\mathrm{k}),
\end{align}
where the index $k$ iterates over all possible erasure combinations on three modes (no erasure, erasure on mode 1, erasure on mode 2, and so forth), $\rho_\mathrm{k}$ is the corresponding state after error correction for combination $k$, and $p_k$ the probability of each combination. In the combinations when error correction is not possible, e.g. erasures on modes 1 and 2,
%($k=1\&2$), 
$\rho_{k}=\tr_\mathrm{M}[\rho^\mathrm{in}_\mathrm{MN}]$. 
Using the upper bounds of the RCI as a new metric, we are interested in comparing the protocol with direct transmission. Direct transmission, in this case, corresponds to transmitting mode $\mathrm{M}$ (of the state $\ket{\psi_\mathrm{MN}}$) via the channel. We compute the upper bounds of the RCI for the simplified channel used for Fig.~\ref{fig:nonG}.

\begin{figure}
   \centering
   \includegraphics[width=.5\textwidth]{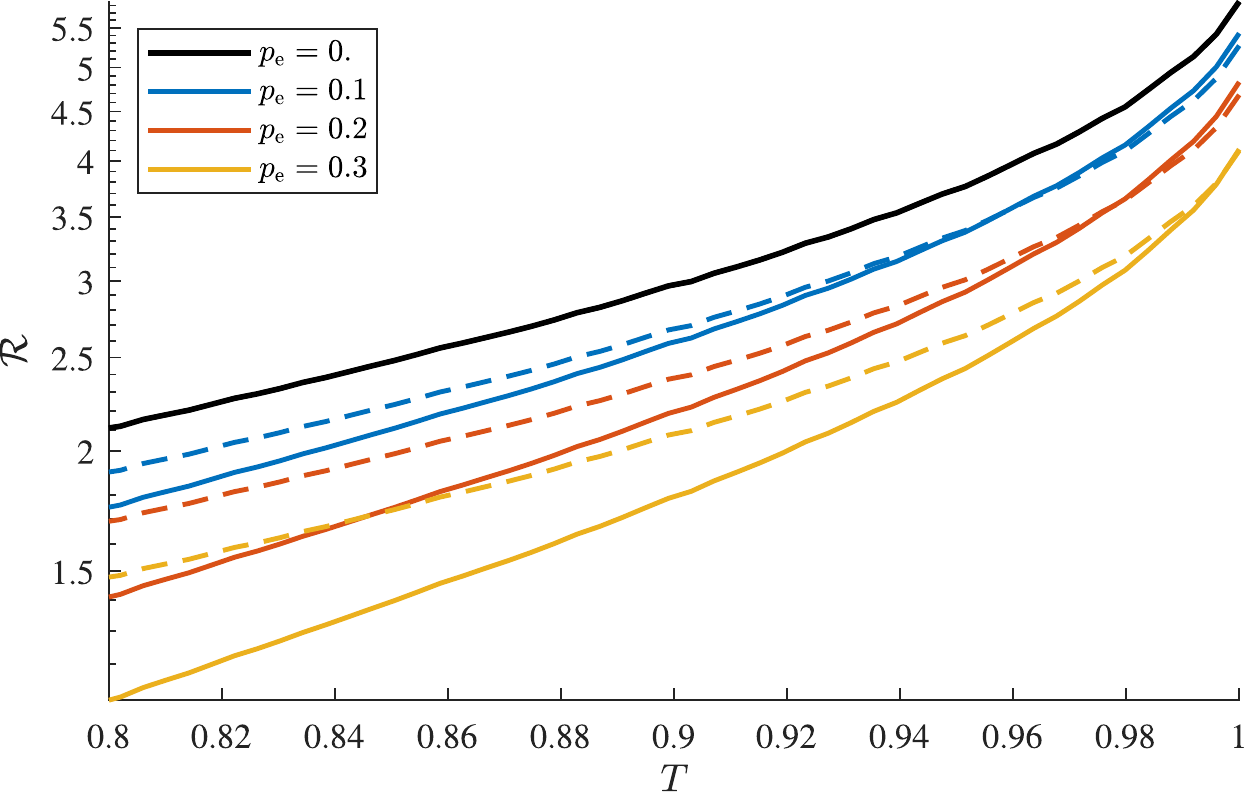}
   \caption{RCI obtained using the protocol (solid lines), and by direct transmission (dashed lines) over a fixed loss channel with a probability of erasure $p_\mathrm{e}$. The values $\epsilon=\epsilon_\mathrm{det}$, and $\eta=1$ are used. A maximally entangled TMSV is considered as $\ket{\psi_\mathrm{AB}}$.}
   \label{fig:RCI}
 \end{figure}

The resulting upper bounds of RCI are presented in Fig.~\ref{fig:RCI}. The results show that the upper bound on the RCI obtained from the protocol and direct transmission is virtually equal for values $T'$ close to 1; as $T'$ decreases in value, the protocol stops presenting an advantage, and the RCI becomes lower than the one obtained by direct transmission.
The lack of an advantage by the protocol in this simulation can be understood by the fact that loss in the ancilla modes (modes $2'$ and $3'$) in the protocol propagates to the final state.
Thus, while the transmission of coherent states can be effectively enhanced by using the protocol presented here, we warn that this may not be the case for applications that are dependent on the distribution of entangled states.

\section{Conclusions}
\label{sec:conclusions}

In this work, a new
 error correction protocol was introduced. Aimed at deployment over realistic free-space optical channels where at least one transceiver is untethered, the protocol was designed to encompass both classical and quantum information
on the encoded signal. We showed, relative to non-encoded
direct transmission, that the protocol improves the fidelity of
transmitted coherent states over a wide range of losses and
erasure probabilities. In addition, using ancillary
non-Gaussian entangled bipartite states in the signal encoding
further increased the fidelity of transmission.
Variants on a theme for the protocol were discussed, including its application to the transmission of components of a multi-mode
entangled state. Our results demonstrate that signal loss in atmospheric channels can be compensated for significantly at the price of additional implementation complexity.

\section*{Acknowledgements}
We thank Ziqing Wang for the valuable discussions. The Australian Government supported this research through the Australian Research Council's Linkage Projects funding scheme (project LP200100601). The views expressed herein are those of the authors and are not necessarily those of the Australian Government or the Australian Research Council. Approved for Public Release; Distribution is Unlimited; \#23-0238.

 \begingroup
 \raggedright

 \bibliographystyle{unsrt}
 \bibliography{references}
 \endgroup

%%%%%%%%%%%%%%%%%%%%%%%%%%%%%%%%%%%%%%%%%%%%%%%%%%%%%%%%%%%%%%%%%%%%%%%
%%%% EQUATION DERIVATION %%%%%%%%%%%%%%%%%%%%%%%%%%%%%%%%%%%%%%%%%%%%%
\begin{appendices}
\section{Output state CF derivation}
\label{sec:ap:derivation}
We will derive the CF of the output quantum state obtained by the protocol. To do so, we follow the diagram presented in Fig.~\ref{fig:qec}.
As the first step, consider the application of a displacement operation on mode 2 of a bipartite state $\hat{\rho}_\mathrm{AB}$,
\begin{align}
  \rho'  = \hat{D}(\Delta) \hat{\rho}_\mathrm{AB} \hat{D}^\dag(\Delta).
\end{align}
Then the CF of the state $\rho'$ is,
\begin{align}
  \bigchi_{\rho'}(\lambda_1, \lambda_2) &= \tr[\hat{D}(\Delta) \hat{\rho}_\mathrm{AB}  \hat{D}^\dag(\Delta) \hat{D}(\lambda_1) \hat{D}(\lambda_2) ]
\end{align}
Using the following properties of the displacement operator,
\begin{align}
  &\hat{D}(\alpha)\hat{D}(\beta) = e^{(\alpha \beta^* - \alpha^* \beta)/2}\hat{D}(\alpha + \beta); \nonumber\\
  &\hat{D}^\dag(\alpha)\hat{D}(\beta)\hat{D}(\alpha) = e^{(\alpha^* \beta - \alpha \beta^*)}\hat{D}(\beta),
\end{align}
the CF of $\hat{\rho}'$ is then,
\begin{align}
  \bigchi_{\hat{\rho}'}(\lambda_1, \lambda_2)&=\tr[ e^{(\Delta^* \lambda_2 - \Delta \lambda_2^*)} \hat{D}(\lambda_1) \hat{D}(\lambda_2)  \hat{\rho}_{AB}]\nonumber \\
  &= \bigchi_\mathrm{AB}(\lambda_1, \lambda_2) \mathrm{D}_{\Delta}(\lambda_2)
\end{align}
where we defined the function  $\mathrm{D}_{\Delta}(\lambda) = e^{(\Delta^* \lambda_ - \Delta \lambda^*)}$ and $\bigchi_\mathrm{AB}$ is the CF of $\hat{\rho}_{AB}$.
Now the CF of the initial state at the start of the protocol, considering the displacement applied to the entangled state, is
\begin{align}
  \bigchi_\mathrm{s}\left(\lambda_1\right)\bigchi_\mathrm{AB}\left(\lambda_2, \lambda_3\right)\mathrm{D}_{\Delta}\left(\lambda_2\right).
\end{align}
The effect of BS1 acting on modes 1 and 3 can be described as a change in variables, such as
\begin{align}
  \bigchi_\mathrm{s}\left(\frac{\lambda_1 + \lambda_2}{\sqrt{2}}\right)
  \bigchi_\mathrm{AB}\left(\frac{\lambda_1-\lambda_2}{\sqrt{2}}, \lambda_3\right) 
 \mathrm{D}_{\Delta} \left(\frac{\lambda_1 -\lambda_2}{\sqrt{2}}\right).
\end{align}
After that, the effect of the loss acting independently on every channel transforms the CF to
\begin{align}
  &\bigchi_\mathrm{s}\left(\frac{\sqrt{T_1}\lambda_1 + \sqrt{T_2}\lambda_2}{\sqrt{2}}\right)
  \bigchi_\mathrm{AB}\left(\frac{\sqrt{T_1}\lambda_1-\sqrt{T_2}\lambda_2}{\sqrt{2}}, \sqrt{T_3}\lambda_3\right) \nonumber \\
  &\times \mathrm{D}_{\Delta} \left(\frac{\sqrt{T_1}\lambda_1 -\sqrt{T_2}\lambda_2}{\sqrt{2}}\right)
   \bigchi_{\ket{0}}\left( T_1^* \lambda_1 \right) \bigchi_{\ket{0}}\left( T_2^* \lambda_2 \right) \nonumber \\  
  &\times \bigchi_{\ket{0}}\left( T_3^* \lambda_3 \right),
\end{align}
with $T_j^*=\sqrt{1 + \epsilon - T_j}$.
Thereafter, BS2 is applied by Bob on modes 1' and 2',
\begin{align}
 &\bigchi_\mathrm{s}\left(\frac{\sqrt{T_1}(\lambda_1 + \lambda_2) + \sqrt{T_2}(\lambda_1 - \lambda_2)}{2}\right)\nonumber \\
  &\times \bigchi_\mathrm{AB}\left(\frac{\sqrt{T_1}(\lambda_1 + \lambda_2) - \sqrt{T_2}(\lambda_1 - \lambda_2)}{2}, \sqrt{T_3}\lambda_3\right) \nonumber\\
  &\times \mathrm{D}_{\Delta} \left(\frac{\sqrt{T_1}(\lambda_1 + \lambda_2) - \sqrt{T_2}(\lambda_1 - \lambda_2)}{2}\right) \nonumber \\
  &\times\bigchi_{\ket{0}}\left( \frac{T_1^* (\lambda_1+\lambda_2)}{\sqrt{2}} \right)
  \bigchi_{\ket{0}}\left( \frac{T_2^* (\lambda_1-\lambda_2)}{\sqrt{2}}\right)
  \bigchi_{\ket{0}}\left( T_3^* \lambda_3 \right).
\end{align}
Finally, to perform a dual homodyne measurement, BS3 is applied, transforming the state.
%Before the homodyne measurements are performed, we account for the efficiency of the homodyne measurements.
The efficiency of the homodyne measurements can be modeled by considering a BS with transmissivity $\eta^2$, with an extra vacuum mode placed before the detectors.
Then the CF of the state  before homodyne measurements is,
\begin{align}
 \bigchi_\mathrm{BS3} (\lambda_1, \lambda_2, \lambda_3)&= \nonumber \\ 
 \bigchi_\mathrm{BS3}' (\lambda_1, \lambda_2,& \lambda_3) 
  \mathrm{D}_{\Delta} \left(\lambda_1 T_{\minus} + \frac{(\lambda_2 + \lambda_3)}{\sqrt{2}}\eta T_{\plus}\right),
    \label{eq:cfbs}
\end{align}
where
\begin{align}
& \bigchi_\mathrm{BS3}' (\lambda_1, \lambda_2, \lambda_3)= \nonumber \\ 
  &~\bigchi_\mathrm{s}\left(\lambda_1 T_{\plus} + \frac{\lambda_2 + \lambda_3}{\sqrt{2}}\eta T_{\minus} \right) \nonumber \times \\
  &~ \bigchi_\mathrm{AB}\left(\lambda_1 T_{\minus} + \frac{(\lambda_2 + \lambda_3)}{\sqrt{2}}\eta T_{\plus} , \frac{\sqrt{T_3} \eta (\lambda_2-\lambda_3)}{\sqrt{2}}\right) \nonumber \times \\
 &~ \bigchi_{\ket{0}}\left( T_1^* \left(\frac{\lambda_1}{\sqrt{2}} +\eta \frac{\lambda_2 + \lambda_3}{2}\right) \right)\nonumber \times \\
 &~ \bigchi_{\ket{0}}\left( T_2^* \left(\frac{\lambda_1}{\sqrt{2}} -\eta \frac{\lambda_2 + \lambda_3}{2}\right)\right) \nonumber \times \\
  &~\bigchi_{\ket{0}}\left( T_3^* \eta \frac{\lambda_2 -\lambda_3}{\sqrt{2}}  \right) \bigchi_{\ket{0}}\left(\sqrt{1-\eta^2} \lambda_1 \right) \nonumber \times \\
  &~  \bigchi_{\ket{0}}\left(\sqrt{1-\eta^2} \lambda_2 \right).
    \label{eq:cfbs2}
\end{align}

It is convenient to express the measurement results in the {\it phase space} representation by transforming the complex arguments into two real numbers, $x_j = \lambda_j + \lambda_j^*$ and $p_j=i(\lambda_j^* - \lambda_j)$.
Then the homodyne measurements are represented by integrating the measured modes.
\begin{align}
  \bigchi_\mathrm{m}(x, p) &= \frac{\mathcal{P}(\tilde{x}, \tilde{p})^{-1}}{(2\pi)^2} \int d x_2 d p_3 \bigchi_\mathrm{BS3}\left(x, p, x_2, 0, 0, p_3 \right) \nonumber \\ &~~~~~~~~~~~~~~~~~~~\times e^{-i\tilde{x}p_3 + i\tilde{p}x_2},
\label{eq:CFm}
\end{align}
with $\tilde{\mathcal{P}}(\tilde{x}, \tilde{p})$ the PDF of any pair of measurement results, given by
\begin{align}
  \tilde{\mathcal{P}}(\tilde{x}, \tilde{p} )= \frac{1}{(2\pi)^2} \int d x_2 d p_3 \bigchi_\mathrm{BS3}\left(0, 0, x_2, 0, 0, p_3 \right) \nonumber \\
  \times e^{-i\tilde{x}p_3 + i\tilde{p}x_2}.
  \label{eq:prob_dist}
\end{align}
Expanding the terms in $\mathrm{D}_{\Delta}$, we have, 
\begin{align}
\mathrm{D}_{\Delta} &\left(x_1 T_{\minus} +  \frac{x_2}{\sqrt{2}}\eta T_{\plus}, p_1 T_{\minus} + \frac{p_3}{\sqrt{2}}\eta T_{\plus} \right) =  \nonumber \\ 
&\exp\Big[-ix_1 T_{\minus}\im{\Delta} +i p_1  T_{\minus}\re{\Delta}\Big]
\times \\ & \exp\Big[-ix_2 \frac{\eta T_{\plus}\im{\Delta}}{\sqrt{2}} +i p_3 \eta \frac{T_{\plus}\re{\Delta}}{\sqrt{2}}\Big]. \nonumber
\end{align}
Inserting this expression into Eq.~\ref{eq:prob_dist} and manipulating the terms, it becomes clear that the mean of the syndrome measurement results is displaced following Eqs. \ref{eq:mu_class}.

The additional exponential term in Eq.~\ref{eq:CFm} indicates that the state after the measurement requires a corrective displacement to be recovered. This corrective displacement must also account for the additional displacement induced by the displacement $\Delta$. The corrective displacement can be implemented via,
\begin{align}
\hat{D}_\mathrm{corr}(\tilde{x}, \tilde{p}) = & \exp \Big[ - ip \left(\sqrt{2} \tilde{x}' g + T_{\minus}\re{\Delta}\right) + \nonumber \\ 
&  i x\left(\sqrt{2} \tilde{p}' g +T_{\minus} \im{\Delta} \right)\Big],
\label{eq:P}
\end{align}
with 
\begin{align}
    \tilde{x}' = \tilde{x}-\mu_{\tilde{x}}, \nonumber \\
    \tilde{p}' = \tilde{p}+\mu_{\tilde{p}},
\end{align}
where $\mu_{\bar{x}(\bar{p})}$ is defined in Eq.~\ref{eq:mu_class}.
Here, the correction includes a factor of $\sqrt{2}$ to compensate for the global factor that appears in the arguments of $\bigchi_\mathrm{BS3}$ in Eq.~\ref{eq:cfbs}.
% Bob's information on which of the modes had an erasure is what will allow him to select the appropriate values of $(g_x,g_p)$.

Finally, since the output CF will depend on a specific set of measurement results,
the mean over all possible measurement outcomes, weighted by their corresponding probability, must be considered, that is
\begin{align}
\bigchi_\mathrm{out}(x, p) &= \int d \tilde{x} d \tilde{p}
\mathcal{P}(\tilde{x}, \tilde{p}) \bigchi_\mathrm{m}(x, p) \hat{D}_\mathrm{corr}(\tilde{x},\tilde{p}) \nonumber \\
&= \frac{1}{(2 \pi)^2} \int d \tilde{x} d \tilde{p} d x_2 d p_3
 \bigchi_\mathrm{BS3} '(x, p,x_2, 0, 0, p_3 ) \nonumber \\ & ~~~~~~~~~~\times e^{i \tilde{x}' (p_3 - \sqrt{2} p g) - i \tilde{p}' (x_2 - \sqrt{2} x g)}.
\end{align}
At this point, the definition of the Dirac delta function, $\frac{1}{2\pi}\iint e^{i\beta x -i\beta \alpha}f(\alpha)d\beta d\alpha = \int \delta (x - \alpha)f(\alpha)d\alpha$, can be used twice to obtain
\begin{align}
  \bigchi_\mathrm{out}(x, p)&= \int d x_2 d p_3 \delta (p_3 - \sqrt{2}p)\delta (x_2 -\sqrt{2}x) \nonumber \\
  & ~~~~~~~~\times \bigchi_\mathrm{BS3}'(x, p, x_2, 0, 0, p_3 ).
\end{align}
Removing the integrands using the properties of the $\delta$ function, the Eq.~\ref{eq:main} presented above is recovered.

\section{Specific examples of error correction}
\label{sec:ap:examples}
To further understand the error correction, we outline the following three cases, corresponding to different values of $g$.
\subsubsection{$g=0$}
\begin{align}
&\bigchi_\mathrm{s}\left(T_{\plus} \lambda \right) 
\bigchi_\mathrm{AB}\left(T_{\minus} \lambda , 0 \right)
\bigchi_{\ket{0}}\left(\sqrt{1 + \epsilon - \frac{T_1 + T_2}{2}} \lambda \right).
\label{eq:g0}
\end{align}
In this case, the resulting state is independent of the value of $T_3$. 
Moreover,
the excess noise in channels $2'$ and $3'$ does not propagate to the output state. However, when $T_1 {\neq} T_2$ imperfect 
destructive interference in BS2 will translate to additional excess noise introduced by the entangled state, which will be proportional to the squeezing of the state. When $T_1=T_2$, the resulting state is equivalent to the one obtained if the signal state were to be transmitted directly through the channel without using the protocol. 
\subsubsection{$g=\frac{1}{\eta}$}
\begin{align}
&\bigchi_\mathrm{s}\left(\sqrt{T_1} \lambda \right) 
\bigchi_\mathrm{AB}\left(\sqrt{T_1} \lambda, \sqrt{T_3} \lambda^* \right) \times \nonumber \\ 
&\bigchi_{\ket{0}}\left(\sqrt{3 + 3\epsilon + 2\left(\frac{1}{\eta^2} -1\right) - 2 T_1 - T_3} \lambda \right).
\label{eq:g1}
\end{align}
The resulting state will be independent of the value of $T_2$. Unlike the previous case, the resulting state is affected by the loss and excess noise affecting modes $2'$ and $3'$. Consider the case when $T_1=T_3$, then it becomes clear that the vacuum contribution appears in the output state three times. On an ideal scenario, with $T_1=T_3=1$, the quantum signal can be fully recovered in the limit $V \rightarrow \infty$ as in this limit $\bigchi_\mathrm{AB}(\lambda, \lambda^*) \rightarrow 1$ (see Eq.~\ref{eq:cftmsv}) \cite{https://doi.org/10.48550/arxiv.2210.10230}.
\subsubsection{$g=-\frac{1}{\eta}$}
\begin{align}
&\bigchi_\mathrm{s}\left(\sqrt{T_2} \lambda \right )
\bigchi_\mathrm{AB}\left(-\sqrt{T_2} \lambda, -\sqrt{T_3} \lambda^* \right) \times \nonumber \\ 
&\bigchi_{\ket{0}}\left(\sqrt{3 + 3\epsilon + 2\left( \frac{1}{\eta^2} -1\right) - 2 T_2 - T_3} \lambda \right).
\label{eq:g-1}
\end{align}
In this case, the resulting state is independent of $T_1$. Similarly to the previous case, the loss and excess noise from the ancillary entangled state propagates to the output state.

\section{Fidelity of Transmission}
\label{sec:ap:expressions}
To compute the fidelity between the states $\ket{\phi_\mathrm{in}}$ and $\ket{\phi_\mathrm{out}}$ first the CF of $\bigchi_\mathrm{out}$ is obtained using Eq.~\ref{eq:main}.  Next, an analytical expression for a single input coherent state can be obtained by solving the integration in Eq.\ref{eq:fidelity}. Finally, this fidelity is averaged over the ensemble of coherent states by solving the integral in Eq.~\ref{eq:avefidelity}.
The expression obtained is,
\begin{align} \label{eq:Fprotocol}
\mathcal{F}= \frac{2}{2\sigma_\alpha\left( T_{\plus} + \tilde{g} T_{\minus} -1\right)^2  + c_1 + 1},
\end{align}
with 
\begin{align}
c_1 =& V\left(\left(T_{\minus} + \tilde{g} T_{\plus}\right)^2 + T_3\tilde{g}^2\right)- \nonumber \\ 
&2\sqrt{V^2-1}\left(T_{\minus} + \tilde{g} T_{\plus}\right)\sqrt{T_3}\tilde{g} + \left(T_{\plus} + \tilde{g} T_{\minus}\right)^2 + \nonumber \\ 
&(1 + \epsilon) (2\tilde{g}^2 + 1) + g^2(1 - \eta^2) - T'.
\end{align}

Additionally, the analytical expression for the fidelity of direct transmission can be obtained by following the same procedure as above, where the CF of the state transmitted directly through the channel replaces $\bigchi_\mathrm{out}$. The expression obtained is,
\begin{align} \label{eq:Fdirect}
\mathcal{\bar{F}}^{\mathrm{DT}}= \frac{2}{2\sigma_\alpha(\sqrt{T}-1)^2 + 2}.
\end{align}
where $T$ is the transmissivity of the channel.

\section{Derivation of the SNR of the measurement results}
\label{sec:ap:snr}
The SNR ratio of the syndrome measurement is used to calculate the bit error rate of the encoded classical communications. To find the SNR, we need to find the variance of the syndrome measurement results, $\sigma_\mathrm{s}$. 
If all the states involved in the protocol are Gaussian, then solving the integrating in Eq.~\ref{eq:prob_dist}, we observe that $\tilde{\mathcal{P}}$ corresponds to a Gaussian distribution, 
\begin{align}
    \tilde{\mathcal{P}}(\tilde{x}, \tilde{p}) = \frac{1}{2\sigma_\mathrm{s}\pi}\exp\left[-\frac{\tilde{x}'^2 + \tilde{p}'^2}{2\sigma_s^2}\right],
\end{align}
where
\begin{align}
\sigma_s^2 = \eta^2 \left( T_{\minus}^2 \sigma_\alpha + c_2 \right) + 1.
\end{align}
with
\begin{align}
c_2 =&  V\left(\frac{T_{\plus}^2}{2} + \frac{T_3}{2}\right) - \sqrt{V^2-1}(T_{\plus}\sqrt{T3}) +  \nonumber \\ 
 &\frac{T_{\minus}^2}{2} + \left(\epsilon- \frac{T'}{2}\right). \nonumber \\
\end{align}

\section{CF of non-Gaussian states}
\label{sec:ap:nonG}
\subsubsection{Photon addition, photon subtraction, and photon catalysis}
In the following, we apply the procedures of \cite{9463774} for determining CFs.
The {\it unnormalized} CF of the state resulting from the application of the non-Gaussian operation to a mode of the TMSV state after it has been transmitted through the  channel is,
\begin{equation}
	\begin{aligned}
		\bigchi'_{\mathrm{PS}}(&\xi_{\mathrm{A}}, \xi_{\mathrm{B}})
		=\frac{T_\kappa-1}{T_\kappa}
		\exp{\left[-\frac{|\xi_{\mathrm{B}}|^2}{2}\right[}\\
		&\times
		\frac{\partial^2}{\partial \xi_{\mathrm{B}} \partial \xi_{\mathrm{B}}^*}
		\bigg[
		\exp{\left[\frac{|\xi_{\mathrm{B}}|^2}{2}\right]}
		f(\xi_{\mathrm{A}}, \xi_{\mathrm{B}}, \sqrt{T_\kappa})
		\bigg],
	\end{aligned}
\end{equation}
where
\begin{equation}
	\begin{aligned}
		&f(\xi_{\mathrm{A}}, \xi_{\mathrm{B}}, \sqrt{T_\kappa})=\int \frac{d \xi^2}{\pi (1-T_\kappa)} \bigchi_\mathrm{TMSV}(\xi_{\mathrm{A}}, \xi)\\
		&\times
		\exp{
			\Big[
			\frac{1+T_\kappa}{2(T_\kappa-1)}(|\xi|^2+|\xi_{\mathrm{B}}|^2)} +
			\frac{\sqrt{T_\kappa}}{T_\kappa-1}
			(\xi_{\mathrm{B}}\xi^*+\xi_{\mathrm{B}}^*\xi)
			\Big].
	\end{aligned}
\end{equation}
Similarly, the unnormalized CF after the PA operator is applied is,
\begin{equation}
	\begin{aligned}
	\bigchi'_{\mathrm{PA}}(\xi_{\mathrm{A}}, & \xi_{\mathrm{B}})
		=(T_\kappa-1)
		\exp{\left[\frac{|\xi_{\mathrm{B}}|^2}{2}\right[}\\
		&\times
		\frac{\partial^2}{\partial \xi_{\mathrm{B}} \partial \xi_{\mathrm{B}}^*}
		\bigg[
		\exp{\left[-\frac{|\xi_{\mathrm{B}}|^2}{2}\right[}
		f(\xi_{\mathrm{A}}, \xi_{\mathrm{B}}, \sqrt{T_\kappa})
		\bigg].
	\end{aligned}
\end{equation}
When the PC operator is applied, the unnormalized CF is,
\begin{equation}
	\begin{aligned}
		\bigchi'_{\mathrm{PC}}(\xi_{\mathrm{A}}, & \xi_{\mathrm{B}})
		=q^2
		\exp{\left[\frac{|\xi_{\mathrm{B}}|^2}{2}\right[}
		\frac{\partial^2}{\partial \xi_{\mathrm{B}} \partial \xi_{\mathrm{B}}^*}
		\bigg\lbrace
		\exp{\left[-|\xi_{\mathrm{B}}|^2\right[}\\
		&\times
		\frac{\partial^2}{\partial \xi_{\mathrm{B}} \partial \xi_{\mathrm{B}}^*}
		\bigg[
		\exp{\left[\frac{|\xi_{\mathrm{B}}|^2}{2}\right[}
		f(\xi_{\mathrm{A}}, \xi_{\mathrm{B}}, \sqrt{T_\kappa})
		\bigg]\bigg\rbrace\\
		&-q
		\exp{\left[\frac{|\xi_{\mathrm{B}}|^2}{2}\right[}
		\frac{\partial}{\partial \xi_{\mathrm{B}}}
		\bigg\lbrace
		\exp{\left[-|\xi_{\mathrm{B}}|^2\right[}\\
		&\times
		\frac{\partial}{\partial \xi_{\mathrm{B}}^*}
		\bigg[
		\exp{\left[\frac{|\xi_{\mathrm{B}}|^2}{2}\right[}
		f(\xi_{\mathrm{A}}, \xi_{\mathrm{B}}, \sqrt{T_\kappa})
		\bigg]\bigg\rbrace\\
		&-q
		\exp{\left[\frac{|\xi_{\mathrm{B}}|^2}{2}\right[}
		\frac{\partial}{\partial \xi_{\mathrm{B}}^*}
		\bigg\lbrace
		\exp{\left[-|\xi_{\mathrm{B}}|^2\right[}\\
		&\times
		\frac{\partial}{\partial \xi_{\mathrm{B}}}
		\bigg[
		\exp{\left[\frac{|\xi_{\mathrm{B}}|^2}{2}\right[}
		f(\xi_{\mathrm{A}}, \xi_{\mathrm{B}}, \sqrt{T_\kappa})
		\bigg]\bigg\rbrace\\
		&+f(\xi_{\mathrm{A}}, \xi_{\mathrm{B}}, \sqrt{T_\kappa}),
	\end{aligned}
\end{equation}
where $q=\frac{T_\kappa-1}{T_\kappa}$.

For the sequential use of the PS and PA operators, the operations PS-PA and PA-PS, we consider the two non-Gaussian operations to use the same beam-splitter transmissivity, $T_\kappa$.
For PS-PA, the unnormalized CF is,
\begin{equation}
	\begin{aligned}
		&\bigchi'_{\mathrm{PS-PA}}(\xi_{\mathrm{A}}, \xi_{\mathrm{B}})\\
		&\quad=q^2
		\exp{\left[\frac{|\xi_{\mathrm{B}}|^2}{2}\right[}
		\frac{\partial^2}{\partial \xi_{\mathrm{B}} \partial \xi_{\mathrm{B}}^*}
		\bigg\lbrace
		\exp{\left[-|\xi_{\mathrm{B}}|^2\right[}\\
		&\quad\times
		\frac{\partial^2}{\partial \xi_{\mathrm{B}} \partial \xi_{\mathrm{B}}^*}
		\bigg[
		\exp{\left[\frac{|\xi_{\mathrm{B}}|^2}{2}\right[}
		f(\xi_{\mathrm{A}}, \xi_{\mathrm{B}}, {T_\kappa})
		\bigg]\bigg\rbrace.\\
	\end{aligned}
\end{equation}
\balance
The unnormalized CF for PA-PS is,
\begin{equation}
	\begin{aligned}
		&\bigchi'_{\mathrm{PA-PS}}(\xi_{\mathrm{A}}, \xi_{\mathrm{B}})\\
		&\quad=(T_\kappa-1)^2
		\exp{\left[-\frac{|\xi_{\mathrm{B}}|^2}{2}\right[}
		\frac{\partial^2}{\partial \xi_{\mathrm{B}} \partial \xi_{\mathrm{B}}^*}
		\bigg\lbrace
		\exp{\left[|\xi_{\mathrm{B}}|^2\right[}\\
		&\quad\times
		\frac{\partial^2}{\partial \xi_{\mathrm{B}} \partial \xi_{\mathrm{B}}^*}
		\bigg[
		\exp{\left[-\frac{|\xi_{\mathrm{B}}|^2}{2}\right[}
		f(\xi_{\mathrm{A}}, \xi_{\mathrm{B}}, {T_\kappa})
		\bigg]\bigg\rbrace.\\
	\end{aligned}
\end{equation}

\subsubsection{Squeezed Bell-like states}
The normalized CF of an SB state is \cite{nonGaussianTeleportation},
\begin{align}
&\bigchi_\mathrm{SB}(\xi_\mathrm{A}, \xi_\mathrm{B}) = (\cos^2(\vartheta) + \sin^2(\vartheta))^{-1/2}
\nonumber \\
&\exp \Big[-\frac{1}{2}\left(|\xi_\mathrm{A}'|^2 + |\xi_\mathrm{B}'|^2 \right) \Big]
  \Big[ \cos^2(\vartheta)  + 2 \cos(\vartheta)\sin(\vartheta)  \nonumber \\
  &\times \Re\{ \xi_\mathrm{A}' \xi_\mathrm{B}'   \} +  \sin^2(\vartheta) (1 - |\xi_\mathrm{A}'|^2) (1 - |\xi_\mathrm{B}'|^2)  \Big],
 \label{eq:sb}
\end{align}
where $\Re\{ z\}$ is the real part of $z$ and the transformation given by Eq.~\ref{eq:Bogo}, is used.

%%%%%%%%%%%%%%%%%%%%%%%%%%%%%%%%%%%%%%%%%%%%%%%%%%%%%%%%%%%%%%%%%%%%%%%%%%%%%%%%
%%%%%%%%%%%%%%%%%%%%%%%%%%%%%%%%%%%%%%%%%%%%%%%%%%%%%%%%%%%%%%%%%%%%%%%%%%%%%%%

\end{appendices}

\end{document}